\documentclass[12pt]{iopart}
\usepackage{graphicx}
\usepackage{psfig}
\usepackage{color}

\begin{document}

\def\dir{figures/}
\def\ra{\rightarrow}

\def\<{\langle}
\def\>{\rangle}
\def\({\left(}
\def\){\right)}
\def\[{\left[}
\def\]{\right]}
\def\Re{\mbox{Re}}
\def\Im{\mbox{Im}}
\def\tr{\mbox{tr}}
\def\Sgn{\mbox{sgn}}
\def\i{{\rm i}}
\def\d{{\rm d}}
\def\e{{\rm e}}

\newcommand{\al}{\alpha}
\newcommand{\be}{\beta}
\newcommand{\de}{\delta}
\newcommand{\De}{\Delta}
\newcommand{\eps}{\epsilon}
\newcommand{\ga}{\gamma}
\newcommand{\om}{\omega}
\newcommand{\Om}{\Omega}
\newcommand{\lam}{\lambda}
\newcommand{\Lam}{\Lambda}

\newcommand{\half}{{\textstyle{1 \over 2}}}

\graphicspath{{final/}}


\title[Form factor of quantum graphs]
{Form factor for large quantum graphs:\\ 
evaluating orbits with time-reversal} 
\author{Gregory Berkolaiko}
\address{Department of Mathematics, University of Strathclyde,
  Glasgow G1 1XH, UK}
\date{\today}

\begin{abstract}
  It has been shown that for a certain special type of quantum graphs
  the random-matrix form factor can be recovered to at least third
  order in the scaled time $\tau$ using periodic-orbit theory.  Two
  types of contributing pairs of orbits were identified, those which
  require time-reversal symmetry and those which do not.  We present a
  new technique of dealing with contribution from the former type of
  orbits.
  
  The technique allows us to derive the third order term of the
  expansion for general graphs.  Although the derivation is rather
  technical, the advantages of the technique are obvious: it makes the
  derivation tractable, it identifies explicitly the orbit
  configurations which give the correct contribution, it is more
  algorithmic and more system-independent, making possible future
  applications of the technique to systems other than quantum graphs.
\end{abstract}
\maketitle

\section{Introduction}

The Bohigas-Giannoni-Schmit conjecture \cite{BGS84} asserts that the
spectral correlations in quantum systems with chaotic classical
analogue fall into several universality classes, depending on the
symmetries of the system.  In particular, the spectral correlations of
systems with time-reversal (TR) symmetry coincide with the relevant
expressions obtained in random matrix theory for Orthogonal Ensembles
of matrices.  This claim was supported by a multitude of numerical
examples \cite{Haake}, but for a long time the only theoretical
advance for individual systems (i.e.  without disorder average) was
Berry's diagonal approximation \cite{Ber85}.  The recent work of
Sieber and Richter \cite{SR01} and Sieber \cite{S02} has renewed the
hope that the universality of spectral correlations can be explained
within periodic-orbit theory.

One of the more convenient statistics is the Fourier transform of the
spectral two-point correlator, the form factor, whose universal
expression for Orthogonal Ensembles is given by the formula
\begin{eqnarray}
  \label{kgoe}
  K_{\rm GOE}(\tau) = 2\tau-\tau\log(1+2\tau)
  = 2\tau-2\tau^{2}+2\tau^{3}+O(\tau^{4})\,,
\end{eqnarray}
when $\tau$ is in the range $0\leq \tau \leq 1$.  For a quantum
chaotic system the form factor can be written in terms of a double sum
over periodic orbits (PO) using the Gutzwiller trace-formula
\cite{Gut71}.  This double sum is the usual starting point for
analysis, in which different classes of periodic orbit pairs are identified
and their contribution is evaluated to reproduce the small-$\tau$
expansion of the prediction (\ref{kgoe}).

Berry \cite{Ber85} calculated the form factor, neglecting all
correlations between POs other than exact symmetries.  Within this
``diagonal approximation'', he obtained the leading order term in
the $\tau$ expansion.  In \cite{SR01,S02} it was shown, that for
uniformly hyperbolic and time-reversal invariant billiards on surfaces
with constant negative curvature the second-order contribution
$-2\tau^{2}$ is related to correlations within pairs of orbits
differing in the orientation of one of the two loops resulting from a
self-intersection of the orbit.  The same result, but without
restriction to uniformly hyperbolic dynamics, was derived for a large
family of quantum graphs \cite{BSW02}.  In particular, some progress
has been made in identifying the exact requirements on the degree of
``chaoticity'' of the graphs.  The further step to third order was performed
in \cite{BSWpre}, where some of the classes of orbits were evaluated
for general graphs.  However for classes of orbits which explicitly
required time-reversal symmetry, the uniform hyperbolicity had to be
assumed.  At this time it became clear that the method used in
\cite{BSW02} would become intractable for the third order in general
graphs and deriving the fourth order term would be quite impossible
even for systems with uniformly hyperbolic dynamics.

The present manuscript presents a different method of dealing with
various classes of orbits.  The underlying idea is the repeated
application of the inclusion-exclusion principle to obtain a
decomposition of orbit pairs into sets of which only a relatively
small proportion gives nonzero contribution.  This technique has
several advantages: the derivation becomes tractable for general
graphs, the orbit classes giving the universal contributions are
identified explicitly, the application of the technique is a
relatively mechanical process which decreases the chance of missing a
contribution and raises hope for a general derivation of the expansion
to all orders.  It is also worth mentioning that the technique is
relatively system-independent in that it does not use any features
specific to quantum graphs, operating on rather abstract {\em
  diagrams}.

This article is organised as follows: In Section~\ref{sec:PO} we
define our model and explain how the form factor can be expressed as a
double sum over periodic orbits.  In Section~\ref{sec:diag} this sum
is rewritten in terms of diagrams, representing all orbits with a
given number and topology of self-intersections.  Diagrams
contributing to the first three orders are identified.  In
Section~\ref{sec:sum} we explain our method by re-deriving the second
order contribution and then proceed to apply it to obtain the third
order term.


\section{Quantum graphs and periodic-orbit theory}
\label{sec:PO}

We consider graphs with $N$ vertices connected by $B$
directed bonds. A bond leading from vertex $m$ to vertex $l$ is
denoted by $(m,l)$.  Since we are considering graphs with
time-reversal invariance, it is necessary that for any bond $(m,l)$
there exists also the reversed bond $(l,m)$. We do not rule out the
possibility of loops, i.e. bonds of the form $(m,m)$, which are
time-reversal invariant.
 
The discrete quantum dynamics on a graph is defined in terms of a
$B\times B$ unitary time-evolution operator $S^{(B)}$ with matrix
elements
\begin{equation}
  \label{smat}
  S^{(B)}_{m'l',lm}=\delta_{l'l}\,\sigma^{(l)}_{m'm}\e^{\i \phi_{ml}}
\end{equation}
describing the transition amplitudes from the directed bond $(m, l)$
to $(l',m')$.  Here the Kronecker delta ensures that a transition is
possible only between joined bonds and $\sigma^{(l)}_{m'm}$ denotes the
{\em vertex-scattering matrix} at vertex $l$.  The phases $\phi_{ml}$
are random variables distributed uniformly in $[0,2\pi]$ and for a
fixed $B$ they define an ensemble of matrices $S^{(B)}$ which is
used for averaging below.  The form factor is defined at integer times
$t=0,1,\dots$ by
\begin{equation}
  \label{ff}
  K^{(B)}(\tau)=B^{-1}\langle|{\rm tr}S^{t}|^{2}\rangle_{\{\phi\}},
\end{equation}
where $\tau$ is the scaled time $\tau=t/B$ and $\langle\cdots\rangle$
denotes the averaging.  We are interested in the limit of large graphs
$B\to\infty$, keeping the scaled time $\tau$ fixed
\begin{equation}
  \label{fflimit}
  K(\tau)=\lim_{B\to\infty} K^{(B)}(\tau)\,,
\end{equation}
since this is equivalent to the semiclassical limit of chaotic
systems \cite{KS97,KS99}.  It is in this limit that the form factor is
expected to assume the corresponding universal form (\ref{kgoe}).

The classical analogue of the quantum graph \cite{KS97,KS99,BG00} is
represented by a Markov chain on the graph, specified by the doubly
stochastic matrix $M$ of transition probabilities,
\begin{eqnarray}
  \label{matrixM}
  M_{m'l,lm} = |S^{(B)}_{m'l,lm}|^2 = |\sigma^{(l)}_{m'm}|^2 \ .
\end{eqnarray}
Matrix elements of powers of $M$ give the classical probability to get
from bond $(m, l)$ to bond $(k, n)$ in $t$ steps
\begin{equation}
  \label{classtransprob}
  P_{(m,l) \to (k,n)}^{(t)}=\left[M^t\right]_{nk,lm}\,.
\end{equation}
Under very general assumptions it can be shown that the dynamics
generated by $M$ is ergodic and mixing \cite{Grim}, i.~e.\ for fixed
$B$ and $t\to\infty$ all transition probabilities become equal
\begin{equation}
  \label{eq:ergodic}
  P_{(m,l) \to (k,n)}^{(t)} \to B^{-1} \quad\mbox{ as } t\to\infty
  \qquad \forall (m,l), (k,n)\,.
\end{equation}
However, since in (\ref{fflimit}) the limits $B\to\infty$ and
$t\to\infty$ are connected by fixing $\tau$, we need a stronger
condition such as
\begin{equation}
  \label{eq:ergodic-tau}
  P_{(m,l) \to (k,n)}^{(\tau B)} \to B^{-1} \quad\mbox{ as } B\to\infty
  \qquad \forall (m,l), (k,n)\,.
\end{equation}
The requirements on the speed of convergence depend on the order to
which agreement with (\ref{kgoe}) is required.  To avoid unpleasant
estimates we will restrict ourselves to graphs for which the
convergence is faster than any power of $B$.

Another, somewhat related, condition is that following a self-retracing orbits
of length $O(B)$ has negligible probability.   More rigorously, define matrix
$R$ by squaring the elements of the matrix $M$,
\begin{eqnarray}
  \label{matrixR}
  R_{m'l,lm} = |S^{(B)}_{m'l,lm}|^4 = |\sigma^{(l)}_{m'm}|^4 \ .
\end{eqnarray}
We require that for any $\gamma$
\begin{equation}
  \label{eq:matrixR_decays}
  \sum_{n,k,l,m} \left[R^{\gamma B}\right]_{nk,lm} \to 0 
  \quad \mbox{ as } B\to\infty.
\end{equation}
faster than any power of $B$.

The above conditions are in agreement with what is known about various
special classes of graphs.  The conditions are satisfied, for example,
by complete graphs with either Neumann or Fourier vertices
\cite{BSWpre} where numerical evidence strongly supports the BGS
conjecture \cite{KS99,Tan01} and are not satisfied by graphs with
Dirichlet vertices and Neumann star graphs \cite{BK99}, which are
known to produce non-RMT statistics.

A connection between the quantum form factor (\ref{ff}) and the
classical dynamics given by (\ref{matrixM}) can be established by
representing the form factor as a sum over (classical) POs.  A PO is a
sequence of vertices $P=[p_{1},\dots,p_{t}]$ defined up to a cyclic
shift and such that $(p_i, p_{i+1})$ is a bond of the graph for every
$i=1, \ldots, t$.

By expanding the matrix powers of $S$ in (\ref{ff}) and averaging over
phases $\phi$ one arrives at the PO expansion of the form factor
\cite{KS99,BSWpre}
\begin{equation}
  \label{eq:ff-PO-TR}
  K^{(B)}_{\rm TR}(\tau)={t^{2}\over B}
  \sum_{P,Q}A_{P}A_{Q}^*  \ ,
\end{equation}
where $A_P$ is the product of $\sigma$-matrices along the orbit $P$:
$A_P = \sigma_{p_1p_3}^{(p_2)}
\sigma_{p_2p_4}^{(p_3)}\cdots\sigma_{p_np_2}^{(p_1)}$, the star
dentoes the complex conjugation and, most importantly, the sum is
taken over only those pairs of orbits that visit {\em the same set of
  bonds, or their time-reverses, the same number of times}.


\section{The expansion in self-intersections of the periodic orbits}
\label{sec:diag}

\subsection{From orbits to diagrams}
\label{sect:orb_diag}

The calculation of the form factor is now reduced to a combinatorial
problem: ensure that $P$ and $Q$ pass through the same non-directed
bonds.  This can be done by composing $P$ and $Q$ from the same
segments, or {\em arcs}, which would appear in $P$ and $Q$ in
different order and/or orientation.  We classify the pairs of orbits
in the following manner.  We fix a {\em transformation}: a permutation of
arcs followed by the time reversal of selected arcs, and then sum over
all orbits $P,Q$ related by this transformation.  The clearest way to
represent a transformation is graphical, hence we refer to them as
{\em diagrams}. The sum over all diagrams finally gives the form factor.

\begin{figure}[t]
  \def\fscale{0.7}
  \centerline{\includegraphics[scale=\fscale]{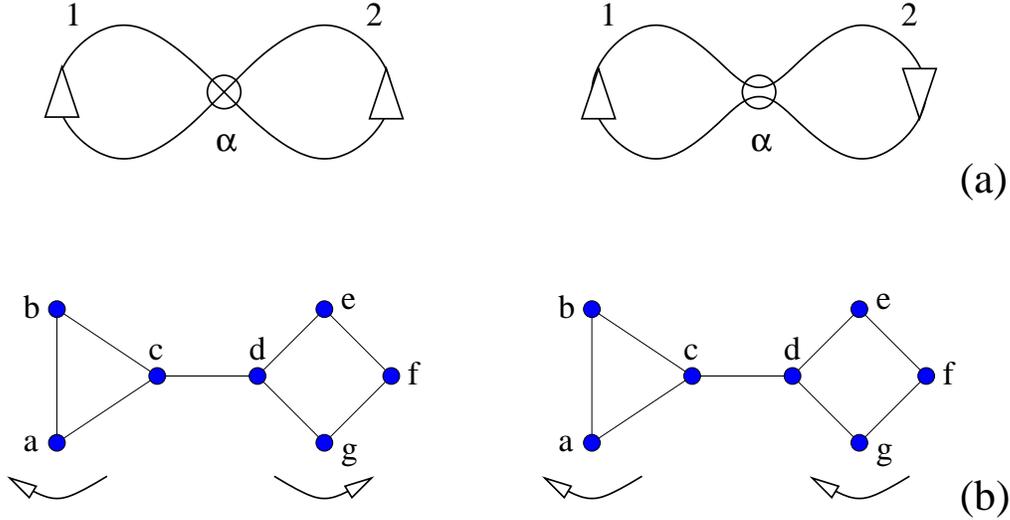}}
  \caption{A schematic representation of an orbit with a
    self-intersection at a vertex $\alpha$ and its partner orbit (a).
    There are orbits, however, for which the position of the
    intersection point $\alpha$ is ambiguous.  The pair of orbits
    shown on (b) can be fitted into the pattern (a) with either $c$ or
    $d$ playing the role of the intersection vertex.}
  \label{fig:TR2}
\end{figure}

The main problem with this approach is to ensure that each orbit pair
$P,Q$ is counted once and only once. This is difficult because for a
given pair $P,Q$ the identification of the arcs and their permutation,
transforming $P$ into $Q$, is not necessarily unique.  As a simple
example consider the diagrams on Fig.~\ref{fig:TR2}.  Part (a) gives a
schematic representation of the pair $P,Q$ where $Q$ is obtained from
$P$ by reversal of arc 2.  Such orbits were considered in \cite{BSW02}
and were shown to give the contribution $-2\tau^2$.  One of the
difficult points of the derivation was correct counting of the orbits
which do not merely cross but follow themselves for at least one bond.
An example of such orbit is given in part (b), where the orbit crosses
itself along the bond $(c, d)$.  For such orbit, there are two
possibilities to identify arc 1: either as $\ra a \ra b \ra $ or as
$\ra c \ra a \ra b \ra c \ra$, thus taking either $c$ or $d$ as the
intersection point.

To avoid this double-counting, \cite{BSW02} imposed a {\em
  restriction\/} on arc 1.  Denoting the first vertex of the arc by
$s_1$ and the last one by $f_1$, it was demanded that $s_1 \neq f_1$.
This ensured the unique choice of the arcs and the intersection point.
In the example above, the valid choice is $\ra a \ra b \ra $.

Unfortunately, even with this restriction, there were orbits that
should not be counted: those {\em exceptions\/} had self-retracing arc
2.  The self-retracing arc did not change under time-reversal and as a
result, the orbit $Q$ was identical to the orbit $P$, forming a pair
which was already counted in the diagonal approximation.  The
contribution of such orbits had to be subtracted explicitly.

In this paper we present a different counting technique, which is
easily extendable to more complicated diagrams, avoids the
introduction of exceptions and explicitly identifies the orbits which
give the generic contribution.  We will first illustrate our approach
by re-deriving the $-2\tau^2$ contribution and then proceed to
calculate the third order correction.  But before we can do it, we
need to introduce the notation and discuss some preliminary matters.

\subsection{Notation}

\begin{figure}
  \def\fscale{0.7}
  \centerline{\includegraphics[scale=\fscale]{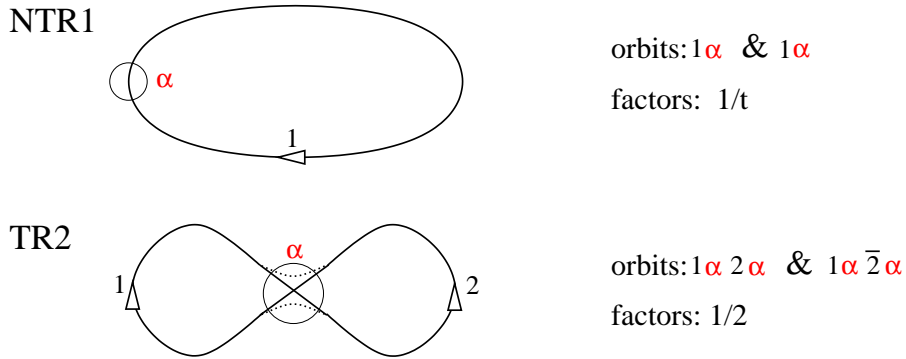}}
  \caption{Topology of orbits contributing to form factor at first and
  second orders.  Evaluating the NTR1 orbits we recover the diagonal
  approximation to the form factor.}
  \label{fig:all12}
\end{figure}

If we consider $P$ as a single arc with no intersections, $Q=P$ and
$Q=\overline{P}$ are the only options.  Here the bar denotes the
operation of time-reversal.  The corresponding diagram have
a simple circular shape (the first diagram of Fig.~\ref{fig:all12}).
Summation over these orbit pairs is nothing other than the diagonal
approximation.  It produces $K_{\rm 1} = 2\tau$.  In \cite{BSW02}
orbit $P$ was treated as two arcs, $1$ and $2$, joined at a single
intersection $\al$, corresponding to an 8-shaped diagram,
Fig.~\ref{fig:all12}.  The contributions of such orbits were found to
give rise to the second-order term in (\ref{kgoe}) $K_{\rm 2} =
-2\tau^{2}$.  In this paper we calculate the $\tau^3$-contribution by
assuming that $P$ contains three or four arcs connected at
intersections.  

We denote arcs by numbers $1, 2, \ldots$ and the intersection points
by Greek letters $\alpha, \beta, \ldots$.  An arc can be identified by
a sequence of vertices, which does not include the intersection
vertices, or, alternatively, by a sequence of bonds, which includes
the bonds from and to the intersection points.  The length of the
$i$th arc is denoted by $t_i$ and is defined as the number of bonds in
the arc (which is one more than the number of vertices in the arc).
The sum of the lengths of all arcs gives $t$, the length of the orbit.
The length of an arc is at least one.  Given an arc $i$ leading from
$\alpha$ to $\beta$ we denote the first vertex following $\alpha$ by
$s_i$ and the last vertex before $\beta$ by $f_i$.  In the degenerate
case when the arc going from $\alpha$ to $\beta$ is the single bond
$(\alpha,\beta)$ and does not contain any vertices ($t_i=1$) our
definition implies $s_i = \beta$ and $f_i = \alpha$.  

\begin{figure}
  \def\fscale{0.7}
  \centerline{\includegraphics[scale=\fscale]{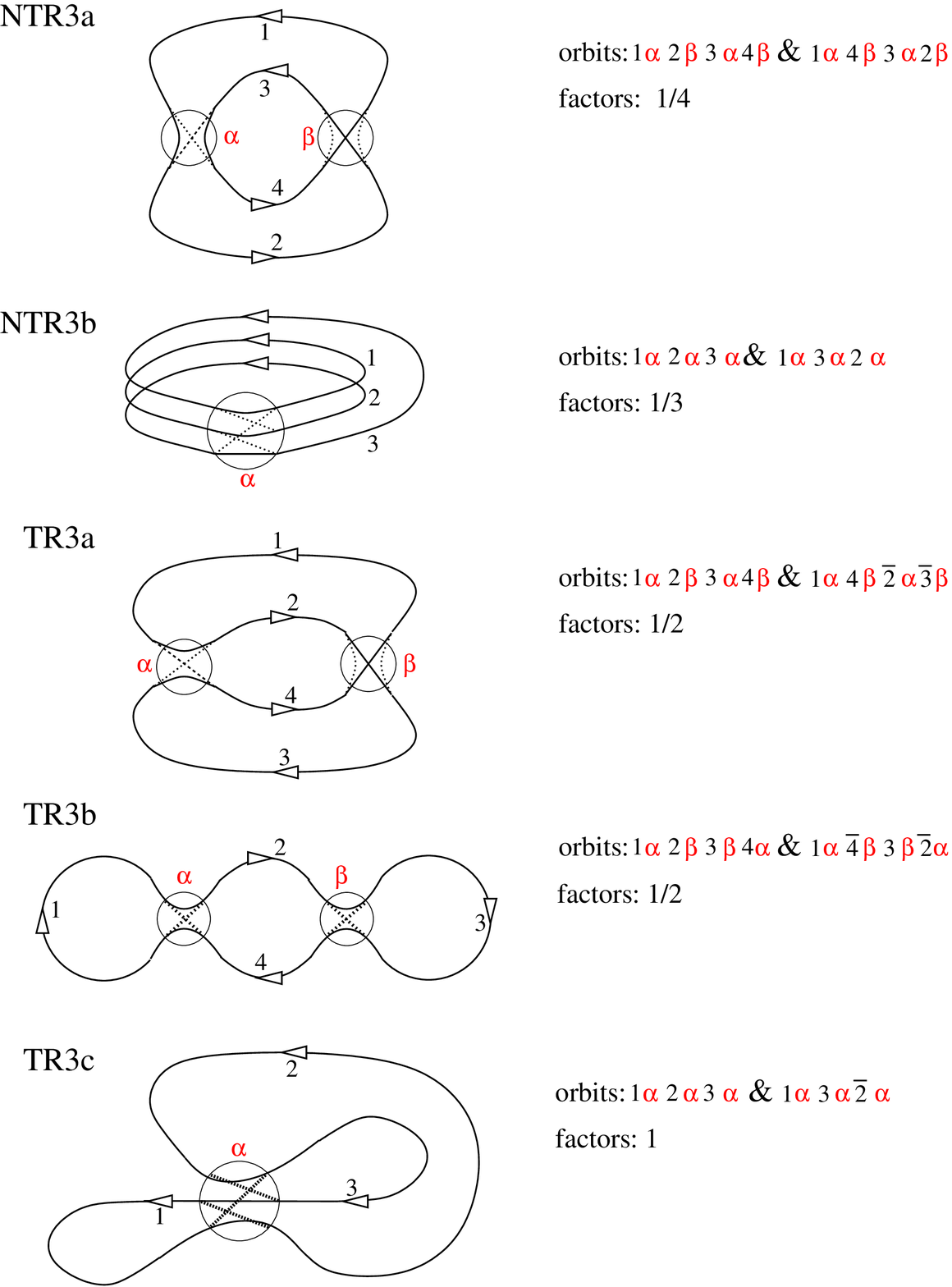}}
  \caption{
    Topology of NTR3a, NTR3b, TR3a, TR3b and TR3c.  In each case a
    pair of orbits is shown, one follows the solid line throughout,
    the second follows the solid line except at the intersections
    (denoted by circles) where it follows the dotted line.  Each
    circle represents a single vertex where a self-intersection of the
    orbit occurs.  Next to each topology we give the symbolic code for
    the pair and the corresponding weight factor
    (Section~\protect\ref{sect:doublecountI}).}  
  \label{fig:all3}
\end{figure}

The diagrams contributing at the third order are shown on
Figs.~\ref{fig:all3}.  For a discussion why only these particular
diagrams contribute at third order in $\tau$ we refer the reader to
\cite{BSWpre}.  In any diagram, the arcs forming orbit $P$ and its
partner $Q$ are identical, but the way they are connected at the
intersections differs.  The orbit $P$ is given by the connections
drawn as continuous lines, while its partner orbit $Q$ is given by
connections drawn as dotted lines.  The orbits $P$ and $Q$ are also
written as a symbolic code to the right of each diagram: a path that
goes from the beginning of arc $1$ to vertex $\al$ then on arc $2$ to
vertex $\be$ and so on is denoted as $1\al 2 \be \cdots$.  The
diagrams divide into two classes, NTR and TR.  In the NTR-diagrams all
the arcs of $Q$ have the same orientation as the corresponding arcs in
$P$, while in the TR-diagrams some of the arcs of $Q$ are
time-reversed.  For a system with no time-reversal symmetry, only the
NTR-diagrams are possible.  In our case, diagrams in both classes
contribute and $\tau^3$-contributions to the form factor is a sum of
five terms
\begin{equation}
  \label{eq:TR3sum}
  K_{\rm 3} = 2 \left( 
    K_{\rm NTR3a}+K_{\rm NTR3b}+K_{\rm TR3a}+K_{\rm TR3b}+K_{\rm TR3c} 
  \right) \  .
\end{equation}
The factor of two is due to the fact that for every diagram in
Fig.~\ref{fig:all3} there is another one with $Q$ replaced by its complete
time-reversal, $\overline{Q}$, which gives an identical contribution. 

The NTR-diagrams were treated in the general case in \cite{BSWpre} and
it was shown that $K_{\rm NTR3a}+K_{\rm NTR3b} = 0$.  In this
manuscript we derive the contribution of TR-diagrams using a slightly
different method.

\subsection{Avoiding double-counting}
\label{sect:doublecountI}
\label{sect:doublecountII}

There are degeneracies in the diagrams which can be accounted for by
simple prefactors multiplying the contributions such as the factor of two in
(\ref{eq:TR3sum}).  These factors were derived in \cite{BSWpre} and
are listed in Fig.~\ref{fig:all3} next to the diagrams.

Another potential source for double-counting of orbits are {\it
  tangential} intersections, already mentioned in
Section~\ref{sect:orb_diag}.  Double-counting can be avoided using a
method outlined in \cite{BSW02} and followed in \cite{BSWpre}: the
intersection point is uniquely defined by ruling that if there is an
ambiguity then the intersection is as far to one side as possible.  As
an example we refer to the ambiguous intersections in
Fig.~\ref{fig:TR2}(b) and insist the intersection is as far to the
left as possible.  That is, we demand that the vertices $a$ and $b$
are distinct.

Contrary to what was done in \cite{BSWpre}, we will not fix the
restrictions for the TR diagrams.  The restrictions we choose will
depend on the lengths of individual arcs, as will be shown in
Section~\ref{sec:sum}.

It should be emphasised that the orbits with tangential
self-intersections are responsible for non-zero contributions to the
form factor and their correct treatment is absolutely crucial to the
derivation.

\subsection{Orbit amplitudes}

Before we can attempt the summation over all orbit pairs $P,Q$ within
a given diagram, we need to understand the structure of the product
$A_P A^*_Q$ appearing in (\ref{eq:ff-PO-TR}).  We consider the diagram
TR2 as an example.  Let arc 1 be of length $t_1$, consisting of the
vertices $[s_1, x_2, x_3, \ldots x_{t_1-2},f_1]$.  Then both $A_P$ and
$A_Q$, will contain factors $\sigma^{(s_1)}_{x_2\,\alpha}$,
$\sigma^{(x_2)}_{x_3x_1}$, $\sigma^{(x_3)}_{x_4x_2} \dots
\sigma^{(x_{t_1-2})}_{\alpha\,x_{t_1-3}}$ .  Thus when we evaluate the
product $A_P A^*_Q$, the contribution of the arc 1 will come in the
form
\begin{equation}
  \label{eq:arc1_contrib}
  |\sigma^{(x_1)}_{x_2\,\alpha} \sigma^{(x_2)}_{x_3x_1}
  \sigma^{(x_3)}_{x_4x_2} \cdots \sigma^{(x_{t_1-2})}_{\alpha\,x_{t_1-3}} |^2 
  = P_{ (\alpha, x_1) \to (x_1, x_2) \to (x_2, x_3)
    \to \ldots \to (f_1, \alpha)}  \equiv P_1,
\end{equation}
which is the classical probability of following arc 1 from bond
$(\alpha,s_1)$ to bond $(f_1,\alpha)$\footnote{ $P_{1}=1$ if arc 1
  contains no vertices, i.~e.\ if $t_1=1$.}.  Analogous construction
leads to the probability $P_2$ of following the arc 2, here we need to
remember that the matrices $\sigma$ are symmetric.  The factors not
yet accounted for in $P_1$ and $P_2$ are the transition amplitudes
picked up at the intersection vertex $\alpha$:
\begin{equation}
  \label{eq:productA_PA_Q}
  A_P A^*_Q = P_1 \times P_2 \times
  \sigma^{(\alpha)}_{f_2s_1} \sigma^{(\alpha)}_{f_1s_2}
  \times
  \Big(\sigma^{(\alpha)}_{s_2s_1} \sigma^{(\alpha)}_{f_1f_2} \Big)^*.
\end{equation}
To evaluate the contribution of the diagram, (\ref{eq:productA_PA_Q})
must be summed over all free parameters, namely all intersection
points and all possible arcs connecting these points. The latter
summation includes a sum over the lengths $t_i$ of these arcs with the
restriction that the total length of the orbit is $t$.

The summation over all the intermediate vertices $x_2, x_3, \ldots,
x_{t_1-2}$ along arc 1 can be performed immediately, since it is
unaffected by the restrictions discussed in the previous subsection.
This summation adds the classical probabilities of all possible paths
leading from bond $(\alpha,s_1)$ to bond $(f_1,\alpha)$ in $t_1-1$
steps and results consequently in the classical transition probability
$P_{(\alpha,s_1) \to (f_1,\alpha)}^{(t_1-1)}$ given by
(\ref{classtransprob}).  Analogous summation over the other arc
produces $P_{(\alpha,s_2) \to (f_2,\alpha)}^{(t_2-1)}$.

The remaining summation is over the lengths $t_i$ of all arcs, the
first and the last vertex $s_i$ and $f_i$ of all arcs $i$ with $t_i>1$
and the intersection points like $\alpha$.  Here we try to use the
fact that for sufficiently long arcs the transition probabilities can
be replaced by $B^{-1}$ according to (\ref{eq:ergodic-tau}).  Then the
sum over vertices decouples into a product of sums associated with the
intersection vertices, which can finally be evaluated using the
unitarity of the vertex-scattering matrices $\sigma$.


\section{Summation of TR diagrams}
\label{sec:sum}

\subsection{TR2}
\label{sec:TR2}
As was mentioned earlier, we develop a slightly different
technique to deal with the TR orbits.  We will illustrate it by first
considering the TR2 contribution.  In \cite{BSW02} we dealt with the
intersection point ambiguity by imposing a restriction $s_2\neq f_2$.
The summation then would take the form
\begin{eqnarray}
  \label{eq:TR2}
  K_{\rm TR2}(\tau) &=& \frac{1}{2}\frac{t^2}{B}
  \sum_{\{t_i\}}\de \left[ t- {\textstyle \sum_{i=1}^2 t_i} \right] \\
  \nonumber
  & & \times \sum_{\alpha} \sum_{\{s_i,f_i\}} 
  \Sigma_{\rm TR2} \times P_{\rm TR2} 
  \times \Delta_{\rm TR2} \ ,
\end{eqnarray}
where 
\begin{eqnarray}
  \label{eq:components_TR2}
  \Sigma_{\rm TR2} 
  &=& \sigma^{\alpha}_{f_2s_1}\sigma^{\alpha}_{f_1s_2}
  \sigma^{\alpha*}_{s_2s_1}\sigma^{\alpha*}_{f_1f_2}\\
  P_{\rm TR2} 
  &=& P_{(\alpha,s_1) \ra (f_1,\alpha)}^{t_1-1}
  P_{(\alpha,s_2) \ra (f_2,\alpha)}^{t_2-1}\\
  \Delta_{\rm TR2} 
  &=& (1-\delta_{s_2f_2}).
\end{eqnarray}

Since at least one of the arcs must be long, we can approximate the
corresponding $P_{(\alpha,s_i) \ra (f_i,\alpha)}^{t_i-1}$ by its
ergodic limit $1/B$.  If arc 1 is long, we can perform the summation
over $s_1$ (and $f_1$)
\begin{equation}
  \label{eq:TR2_zero}
  (1-\delta_{s_2f_2}) 
  \sum_{s_1} \sigma^{\alpha}_{f_2s_1}\sigma^{\alpha*}_{s_2s_1} = 
  (1-\delta_{s_2f_2}) \delta_{s_2f_2} \equiv 0
\end{equation}
to show that the contribution of such orbits is zero.  However, the
case when arc 2 is long is not nearly as easy.  We can still sum over
$s_2$ or $f_2$, but the restriction, which helped us in the first
part, now stands in the way (see \cite{BSW02} for details).

The idea that we are going to use is as follows: we can change the
restrictions depending on the arc lengths.  In TR2 case, if arc 1 is
long, we stick with the restriction $s_2\neq f_2$, and if arc 2 is
long, we switch to the restriction $s_1\neq f_1$.  In the first case
the result of the summation is 0, as was shown above.  In the second case
we approximate $P_{(\alpha,s_2)\ra(f_2,\alpha)}^{t_2-1}$ by $1/B$ and
perform the summation over $s_2$ to obtain
\begin{equation}
  \label{eq:TR2_arc2_long}
  (1-\delta_{s_1f_1}) 
  \sum_{s_1} \sigma^{\alpha}_{f_1s_2}\sigma^{\alpha*}_{s_2s_1} = 
  (1-\delta_{s_1f_1}) \delta_{s_1f_1} \equiv 0.
\end{equation}
This is not what we should be getting: the overall result should be
$-2\tau^2$, not $0$.  The reason for getting $0$ is that there are
orbits which were counted in both sums.  To explain this, we need to
define more carefully what is meant by ``arc $i$ is long''.

\begin{figure}
  \input{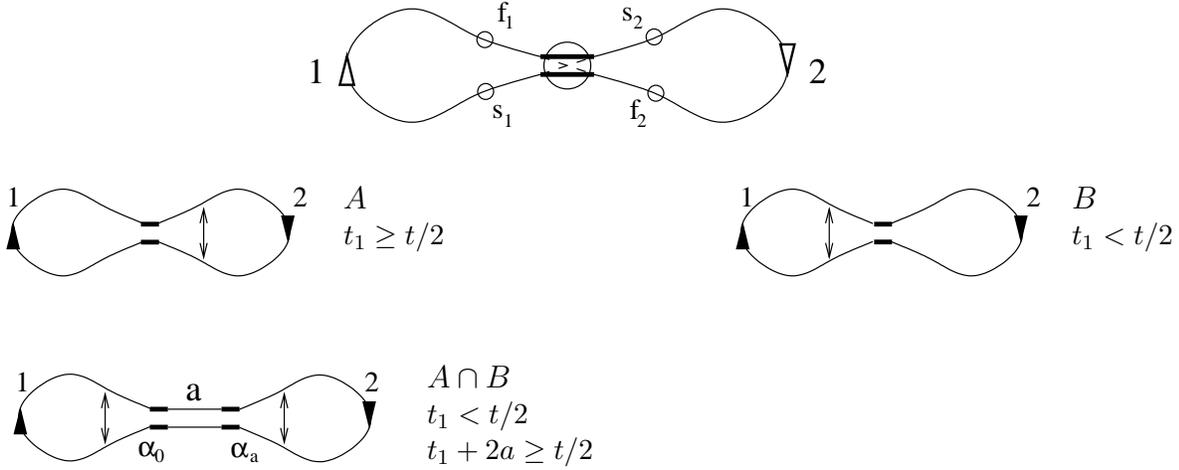}    
  \caption{The TR2 diagram and its partition into two sets of orbits.
  The total contribution of the orbits in each set is easily shown to
  be zero.  The result comes from the orbits lying in the intersection
  of the two sets, $A\cap B$.}
  \label{fig:sets_TR2}
\end{figure}

We split the set of all TR2 orbit pairs into two sets.  The set $A$
will contain orbits satisfying $t_1 \geq t/2$ and $s_2 \neq f_2$.  The
set $B$ will contain orbits satisfying $t_1 < t/2$ and $s_1 \neq f_1$.
The sum of contributions coming from each set is zero, as was shown in
the previous paragraph.  However, these sets are not disjoint.
Fig.~\ref{fig:sets_TR2} illustrates both sets, denoted by $A$ and $B$,
and the shape of the orbits belonging to their intersection $A\cap B$.
These are orbits with a tangential intersection which is roughly in
the middle of them: $t_1 < t/2$ but $t_1 + 2a \geq t/2$, where $a$ is
the length of the intersection and $t_1$ is now counted from the end
of the tangential intersection (i.e. from vertex $\alpha_0$).  We need
to subtract the contribution of such orbits to recover the full term
$K_{\rm TR2}(\tau)$.

To calculate the contribution of the orbits in $A\cap B$, we notice
(see \ref{sec:app1}) that we get the same contribution if we modify
the set $A\cap B$ by setting $a=1$ and dropping the constraints.  Thus
we can write for this contribution
\begin{equation}
  \label{eq:TR2_a_1}
  | A \cap B | =
  \frac{t^2}{2B}
  \sum_{t_1, t_2} \sum_{\alpha_0, \alpha_1} 
  P^{t_1+1}_{(\alpha_1,\alpha_0)\ra(\alpha_0,\alpha_1)}
  P^{t_2+1}_{(\alpha_0,\alpha_1)\ra(\alpha_1,\alpha_0)}
\end{equation}
where $t = t_1 + t_2 + 2$ and, because of the tangential intersection,
the unitary factors at the intersection point got squared and absorbed
into the classical probability.  In fact, the whole expression is
purely classical now.  The length of the arc 1 must satisfy $t_1<t/2$
and $t_1+2\geq t/2$, thus there are only two possible values that
$t_1$ can take, $t/2 - 1$ and $t/2 - 2$.  The length of the arc 2
satisfies $t_2 = t - 2 - t_1$, therefore both arcs are long and the
corresponding probabilities can be approximated by $1/B$,
\begin{equation}
  \label{eq:TR2_a_1_eval}
  | A \cap B | =
  \frac{t^2}{2B^3} \sum_{t_1, t_2} \sum_{\alpha_0, \alpha_1} 1 
  = \frac{t^2}{2B^3} 2 \sum_{\alpha_0, \alpha_1} 1 
  = \frac{t^2}{B^2} \to \tau^2,
\end{equation}
where the last sum is taken over all {\em connected \/} pairs of
vertices $\alpha_0$ and $\alpha_1$ and the factor of 2 appeared
because there are only two possible choices of the lengths $\{t_1,
t_2\}$.  To get the final result for $K_{\rm TR2}(\tau)$, we need to
multiply (\ref{eq:TR2_a_1_eval}) by two once more, to account for the
time-reversal symmetry, and subtract it from the previous result (zero)
since these orbits were double-counted when we calculated the
contributions of the sets $A$ and $B$.  Hence we obtain the
sought-after result $K_{\rm TR2}(\tau) = -2\tau^2$.  It is important
to note that in the above derivation we could have used any $O(t)$
value instead of $t/2$.

Now our strategy of dealing with the diagrams of higher order is
clear.  We will strive to partition the orbits belonging to each diagram
into several sets, such that the contribution of each set is easily
seen to be zero.  The intersections of these set will provide the true
contributions to the form factor.  Hopefully, the arcs of the orbits
in the intersections will be long and thus the ergodic approximation
can be employed to simplify the calculations.

\subsection{Partition of TR3a}

Before we start making partitions of the diagrams, we stress that
there is no unique ``right'' way to partition.  The partition that we
are presenting here is only one of those that we considered.  All
partitions produce the same results, as they should, but the present
one seemed to be the least difficult.

\def\fscale{0.6}
\begin{figure}
  \begin{center}  
    \input{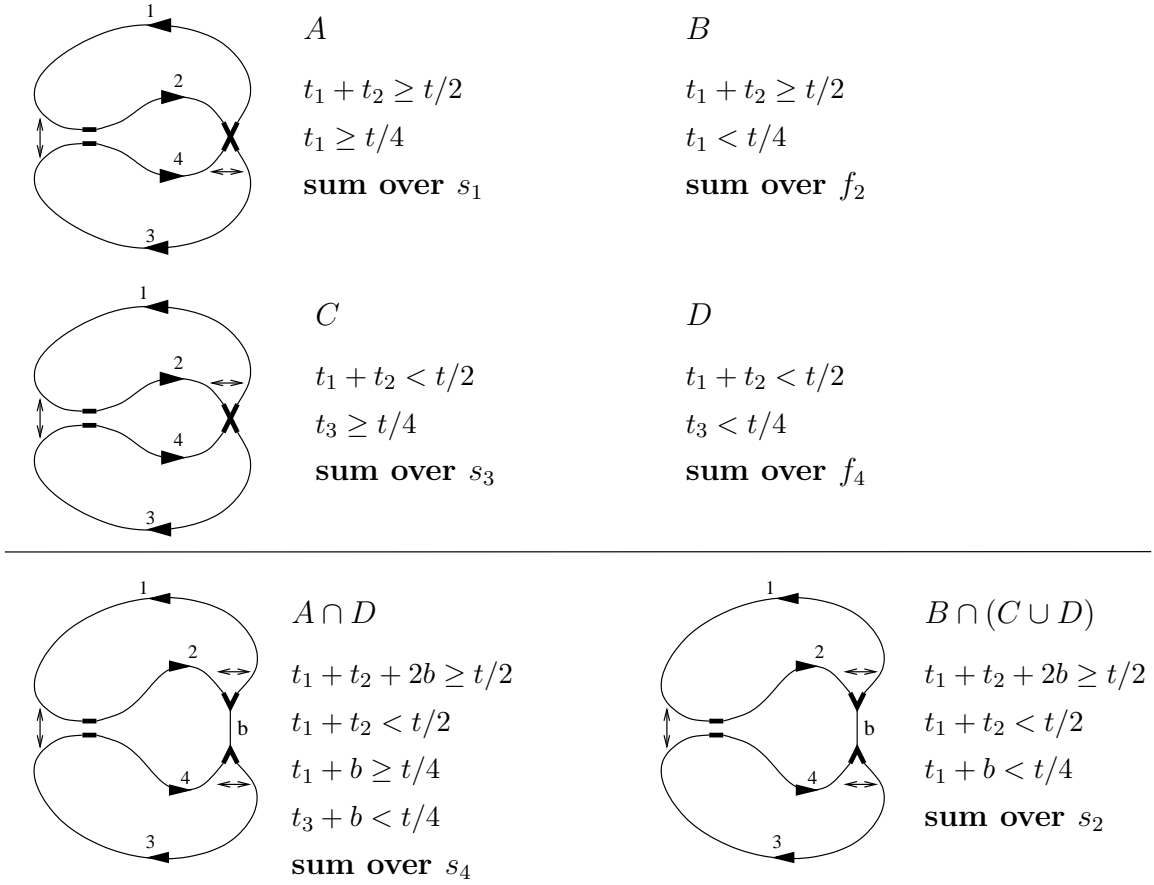}
  \end{center}
  \caption{The TR3a diagram and its partition into four sets of
    orbits.  The orbits of the sets $A$ and $B$ have identical
    structure but different lengths.  The same applies to the sets $C$
    and $D$.  A consequence of this is that $A\cap B = C\cap D =
    \emptyset$ which implies that all 3-way intersections are empty.
    The contribution of the orbits in each set can be easily shown to
    be zero.  Also drawn on the Figure are three of the four 2-way
    intersections among the partition sets of the TR3a diagram.  The
    contribution of each is zero.}
  \label{fig:part_TR3a}
  \label{fig:part_TR3a1}
\end{figure}

The suggested partition for the diagram TR3a is given by
Fig.~\ref{fig:part_TR3a} (upper part).  Each of the 4 subsets of TR3a
is equipped with its own set of constraints, which are shown by the
arrows.  For example, the set $A$ has constraints $f_1\neq f_3$ and
$s_3\neq f_4$.  The constraints are chosen in such a way that the
contributions of each set evaluate to zero: for example, the arc $1$
is long in the set $A$, so we can approximate the contribution of the
arc by it's ergodic limit and subsequently sum over $s_1$ to obtain
zero.  The constraints also take into account the symmetry of the
diagram: under the permutation of arcs $1 \leftrightarrow 3$ and
$2 \leftrightarrow 4$, the set $A$ is interchanged with $C$ and $B$
is interchanged with $D$.  Also, it is immediately obvious that the
sets $A$ and $B$ are disjoint; so are the sets $C$ and $D$.  Therefore
we do not have to worry about 3-way intersections of the partition
sets.

Three of the 2-way intersections are straightforward to evaluate, see
Fig.~\ref{fig:part_TR3a1}.  The intersection $A\cap D$ has $t_1+t_2
\approx t/2$, therefore $t_3+t_4 \approx t/2$ but $t_3+b < t/4$,
therefore the arc $4$ is long and we can sum over $s_4$ obtaining
zero.  The intersections $B\cap C$ and $B\cap D$ can be considered
together, as $B\cap (C\cup D)$.  For this set we again have
$t_1+t_2\approx t/2$ and $t_1$ is small, thus enabling us to sum over
$s_2$.

\begin{figure}
  \begin{center}
    \input{sets_TR3a2.pstex_t}
  \end{center}
  \caption{The last of the 2-way intersections among the partition
    sets of the TR3a diagram.  To evaluate it we represent it as the set
    $E_1$ plus the orbits in the set $E_2$ minus the orbits of the set
    $E_3$.  The contribution of the set $E_1$ is zero.}
  \label{fig:part_TR3a2}
\end{figure}

The last intersection, $A\cap C \equiv E$ is the hardest, see
Fig.~\ref{fig:part_TR3a2}.  Arcs $2$ and $4$ are potentially short and
summation over any other arc is obstructed by the constraints.  To
overcome this, we change the constraint $f_1\neq f_3$ to $s_2\neq
s_4$, thus producing the set $E_1$.  There are certain orbits,
however, that belong to $E$ and do not belong to $E_1$ and vice versa.
These are put into the sets $E_2$ and $E_3$, correspondingly.  The
orbits from the set $E_1$ can be easily shown to produce zero
contribution.  The contribution of the sets $E_2$ and $E_3$ are not
zero and will be evaluated separately.

\subsection{Partition of TR3b}

The diagram TR3b is split into three sets, Fig.~\ref{fig:part_TR3b}.
Here $r$ is a large number, of order $t/4$.  The reason we do not just
put it equal to $t/4$ is to highlight the interaction of the diagram
TR3b with TR3c.

To show that contribution of the sets $A$ and $B$ is zero, simply sum
over $s_1$ and $s_3$, correspondingly.  In set $C$, at least one of
the arcs $2$ and $4$ is long, thus enabling us to sum over $s_2$ or
$s_4$.  It is easy to see that the sets $A$ and $B$ are disjoint.

\begin{figure}
  \begin{center}
    \input{sets_TR3b.pstex_t}
  \end{center}
  \caption{The TR3b diagram and its partition into three sets of orbits.
    The contribution of the orbits in each set can be easily shown to
    be zero.  Below the line we sketch the intersection of the sets
    $A$ and $C$, re-partitioned as $F_1$ minus $F_2$ and the sets $B$
    and $C$, re-partitioned as $F_3$ minus $F_4$ and $F_5$.}
  \label{fig:part_TR3b}
\end{figure}

The intersection $A\cap C$ is also shown on Fig.~\ref{fig:part_TR3b}.
One of the arcs $2$ and $4$ must be long.  To be able to sum, we need
to drop the restriction $f_2 \neq s_4$, producing the set $F_1$.  By
summing either over $f_2$ or $s_4$ (depending on which arc is long),
we show that the contribution of $F_1$ is zero.  However, the set
$F_1$ is larger than $A\cap C$: it includes the orbits sketched in the
sets $F_2$, those can be obtained if, say, arc $4$ is one bond long
and $f_2=s_4$.  The set $F_2$ provides a non-zero contribution and
will be evaluated separately.

In a similar fashion we treat the intersection $B\cap C$, which has to
be split into the sets $F_3$, $F_4$ and $F_5$.  The contribution of
the set $F_3$ is zero and the sets $F_4$ and $F_5$ will be treated
separately.

\subsection{Partition of TR3c}

The diagram TR3c is very special as the orbits belonging to it can
also be obtained as orbits from TR3a or/and TR3b when one of the $t_i$
is equal to zero.  Our job is to synchronise the partition of TR3c
with the partitions we chose for TR3a and TR3b.  The relations between
TR3c and other diagrams are summarised in Table~\ref{tab:TR3c}.

\begin{table}[t]
  \begin{tabular}{|c|c|clll|}
    \hline
    TR3c & TR3a & \multicolumn{4}{c|}{Lengths} \\
    \hline
    $f_2=f_1$ & $s_2=s_4$ ($t_2'=1$)& 
    $t_1' = t_1-1$,\quad & $t_2' = 1$,\quad 
    & $t_3' = t_2-1$,\quad & $t_4'=t_3+1$\\
    $s_2=f_3$ & $s_1=f_2$ ($t_2'=1$)& 
    $t_1' = t_1+1$,\quad & $t_2' = 1$,\quad 
    & $t_3' = t_2-1$,\quad & $t_4'=t_3-1$\\
    $s_1=f_2$ & $s_3=f_4$ ($t_3'=1$)& 
    $t_1' = t_1-1$,\quad & $t_2' = t_2-1$,\quad 
    & $t_3' = 1$,\quad & $t_4'=t_3+1$\\
    $s_2=s_3$ & $f_1=f_3$ ($t_3'=1$)& 
    $t_1' = t_1+1$,\quad & $t_2' = t_2-1$,\quad 
    & $t_3' = 1$,\quad & $t_4'=t_3-1$\\
    $s_1=f_2$ R & $s_1=f_2$ ($t_1'=1$) &
    $t_1'=1$ \quad & $t_2'=t_3+1$ \quad & 
    $t_3' = t_1-1$ \quad & $t_4'=t_2-1$\\
    $s_2=s_3$ R & $f_1=f_3$ ($t_1'=1$) &
    $t_1'=1$ \quad & $t_2'=t_3-1$ \quad & 
    $t_3' = t_1+1$ \quad & $t_4'=t_2-1$\\
    $s_2=f_3$ R & $s_3=f_4$ ($t_4'=1$) &
    $t_1'=t_2-1$ \quad & $t_2'=t_3-1$ \quad & 
    $t_3' = t_1+1$ \quad & $t_4'=1$\\
    $f_1=f_2$ R & $s_2=s_4$ ($t_4'=1$) &
    $t_1'=t_2-1$ \quad & $t_2'=t_3+1$ \quad & 
    $t_3' = t_1-1$ \quad & $t_4'=1$\\
    \hline
    TR3c & TR3b &\multicolumn{4}{c|}{Lengths} \\
    \hline
    $s_1=f_1$ & $s_2=f_4$ 
    & $t_1'=t_1-2$ & $t_2'=t_2+1$ & $t_3'=t_3$ & $t_4'=1$\\ 
    & $s_4=f_2$ 
    & $t_1'=t_3$ & $t_2'=1$ & $t_3'=t_1-2$ & $t_4'=t_2+1$\\
    $s_3=f_3$ & $s_4=f_2$
    & $t_1'=t_1$ & $t_2'=t_2+1$ & $t_3'=t_3-2$ & $t_4'=1$\\
    & $s_2=f_4$
    & $t_1'=t_3-2$ & $t_2'=1$ & $t_3'=t_1$ & $t_4'=t_2+1$\\
    \hline
  \end{tabular}
  \caption{Summary of conversions between degenerate orbits of TR3c and
    two other diagrams.  Here $t_i$ denotes the length of $i$th arc in
    the diagram TR3c and $t_i'$ denotes the same but for the diagrams
    TR3a and TR3b; an `R' denotes that the diagram is reversed.  For
    example, the TR3c orbit with $f_1=f_2$ can be also considered as a
    TR3a orbit with $s_2=s_4$ and lengths of the arcs given by $t_1' =
    t_1-1$, $t_2' = 1$, $t_3' = t_2-1$ and $t_4'=t_3+1$}
  \label{tab:TR3c}
\end{table}

\begin{figure}
  \begin{center}
    \input{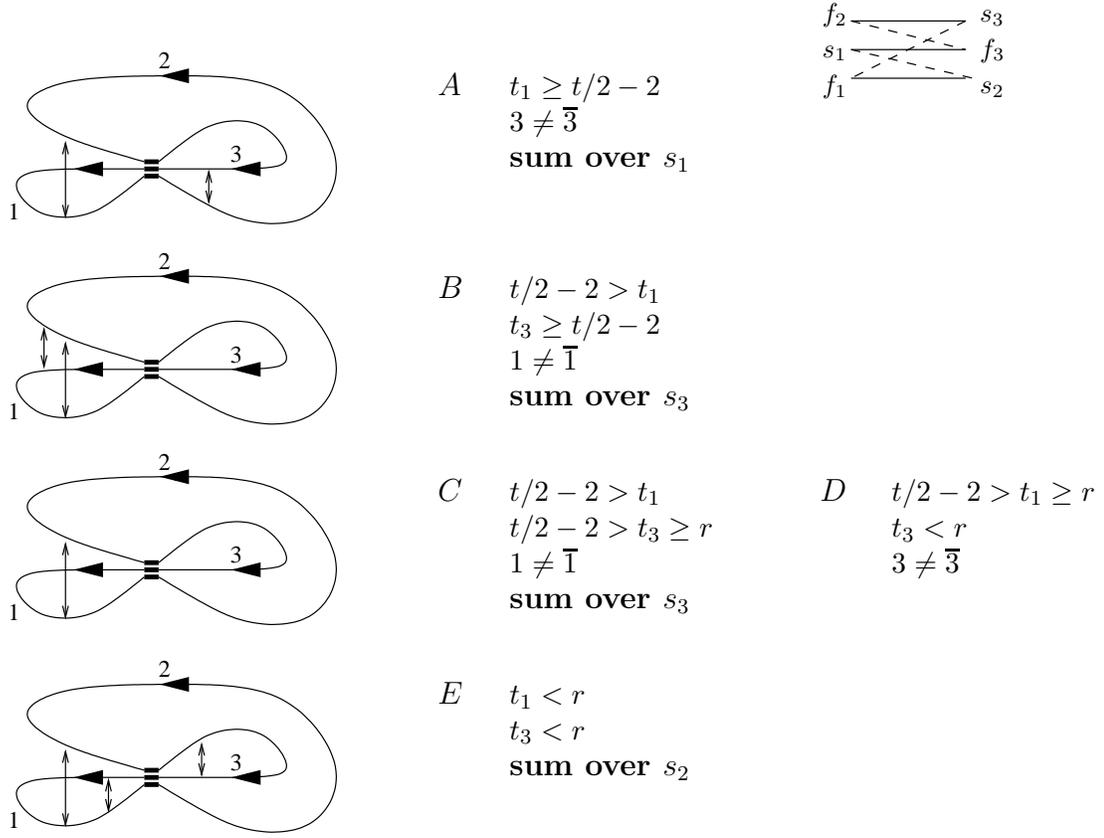}
  \end{center}
  \caption{The TR3c diagram and its partition into five sets of orbits.
    In the upper right corner we illustrate the transitions through
    the intersection vertex undertaken by the original (solid lines)
    and the partner (dashed lines) orbits.  The contribution of all
    sets but $D$ can be easily shown to be zero.  The set $D$ is
    re-partitioned separately in Fig.~\ref{fig:part_TR3c1}.}
  \label{fig:part_TR3c}
\end{figure}

An immediate consequence is that all TR3c orbits with $f_1=f_2$ were
already counted in TR3a: no set in TR3a has the restriction $s_2\neq
s_4$.  Also {\em some\/} of the orbits from TR3c with $s_2=f_3$ are
counted in the sets $A$ and $C$ of TR3a.  To understand precisely
which orbits were counted, we notice that a TR3c orbit with $s_2=f_3$
corresponds to a TR3a orbit with $s_1=f_2$ and $t_2'=1$.  Such orbit
can only belong to the set $A$, requiring $t_1'+t_2' \geq t/2$ and
$t_1'\geq t/4$, which is equivalent to $t_1 \geq t/2-2$ (see the
``Lengths'' column of the table).  Same applies to the reversed orbits
from TR3c with $s_2=f_3$ which were counted in TR3a set $C$.  Thus
orbits from TR3c with $t_1 \geq t/2-2$ must have the restriction
$s_2\neq f_3$ (set $A$ of Fig.~\ref{fig:part_TR3c}).  We also note
that orbits from TR3c with $s_1=f_2$ and $t_3\geq t/2-2$ were already
counted in TR3a sets $B$ and $D$, warranting the use of corresponding
restriction for the set $B$ of Fig.~\ref{fig:part_TR3c}.

The relation between TR3c and TR3b is simpler: if the lengths of an
orbit from TR3c satisfy $t_1<r$ and $t_3<r$, then we must impose
restrictions $s_1\neq f_1$ and $s_3\neq f_3$, which is reflected in
the set $E$ of Fig.~\ref{fig:part_TR3c}.

Bringing the above information together we fix the partition of the
diagram TR3c, Fig.~\ref{fig:part_TR3c}.  All five partition sets are
disjoint and the contributions of the sets $A$, $B$, $C$ and $E$
evaluate to zero by summing with respect to $s_1$, $s_3$ and $s_3$ and
$s_2$ correspondingly.  The set $A$ is slightly special since when
$t_3 \geq t/2-2$, there is an additional restriction $s_1\neq f_2$, but
in this case one can sum over $s_3$ to obtain the zero result.

\begin{figure}
  \input{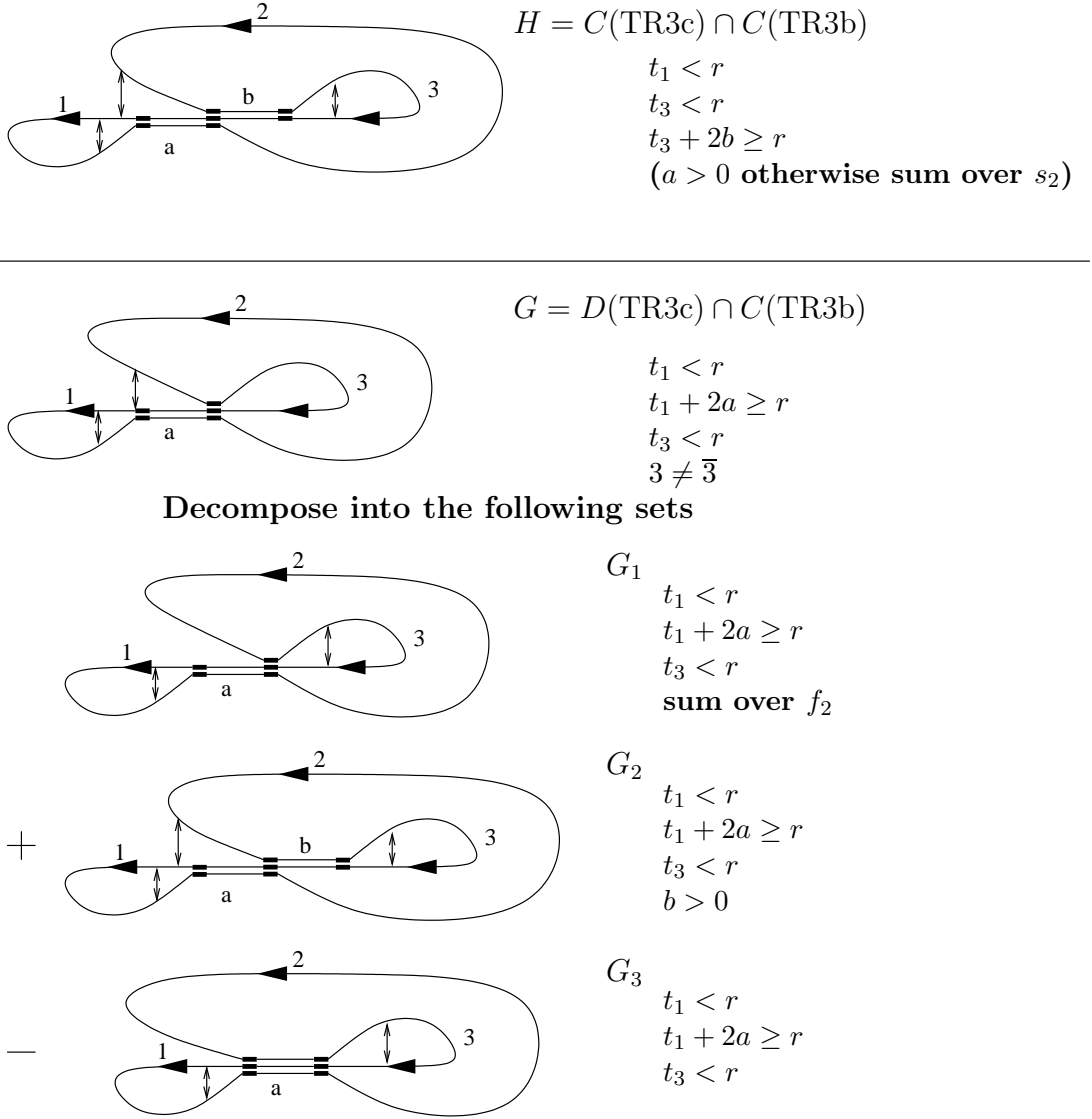}
  \caption{The set of orbits belonging to both set $D$ of TR3c and set
    $C$ of TR3b is re-partitioned as the set $G_1$ plus $G_2$ minus
    $G_3$.  The contribution of the set $G_1$ is zero.}
  \label{fig:part_TR3bc}
\end{figure}

Despite the length restriction on the sets $C$ and $D$, there are
still orbits belonging to these sets and the set $C$ of TR3b.  These
orbits are sketched on Fig.~\ref{fig:part_TR3bc}.  When considering
the intersection of $C$ and TR3b, we assume $a > 0$ since when $a =
0$, we can sum over $s_2$ to get zero.  The contributions of this set
will be evaluated separately.

The intersection of $D$ and TR3b can be re-partitioned as $G_1$ plus
$G_2$ minus $G_3$ with the set $G_1$ giving zero contribution (sum
over $f_2$).  We will show that the contributions of the sets $G_2$
and $G_3$ cancel each other exactly.

\begin{figure}
  \input{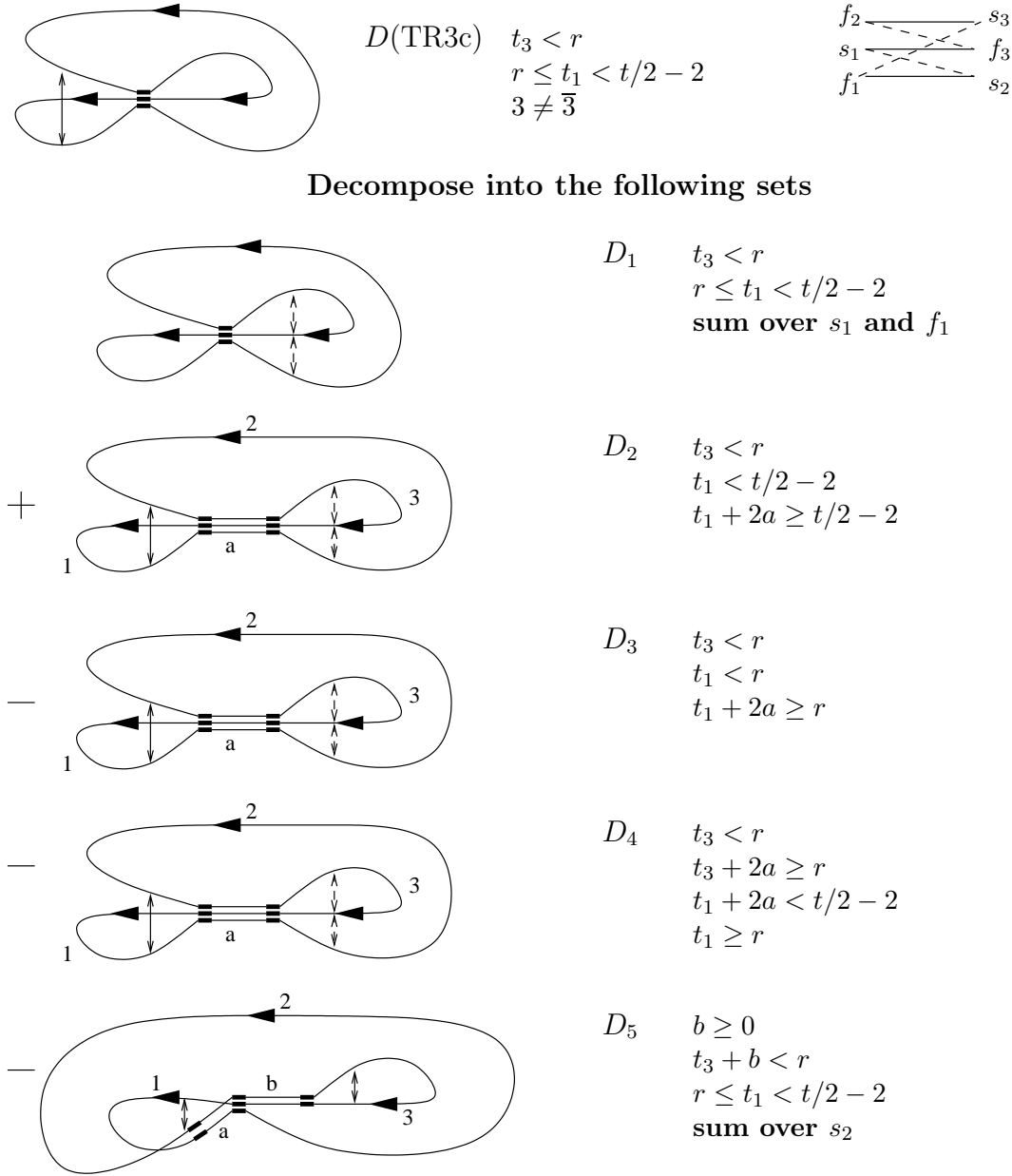}
  \caption{Re-partitioned set $D$ of the TR3c diagram.  Two dashed
    restrictions are denoting the situation when $s_3=f_3$ and
    $s_2=f_3$ cannot be satisfied {\em simultaneously\/}.  The sets
    $D_1$ and $D_5$ give zero contribution and the sets $D_2$ and
    $D_3$ produce contributions that cancel each other.}
  \label{fig:part_TR3c1}
\end{figure}

The last set to be decomposed is the set $D$ of TR3c, see
Fig.~\ref{fig:part_TR3c1}.  Two dashed restrictions are denoting the
situation when $s_3=f_3$ and $s_2=f_3$ cannot be satisfied {\em
  simultaneously\/}.  Out of five partition sets, the sets $D_1$ and
$D_5$ are immediately evaluated to zero, the contributions of the sets
$D_2$ and $D_3$ clearly cancel each other, thus only the contribution
of the set $D_4$ survives.

\subsection{Evaluating significant diagrams}

If we denote the contribution of a set $A$ by $|A|$, the total
contribution to the $\tau^3$ term of the form factor expansion takes
the form
\begin{eqnarray}
  \label{eq:significant_sets}
  &2&(K_{\rm TR3a}(\tau) + K_{\rm TR3b}(\tau) + K_{\rm TR3c}(\tau))
  \\ \nonumber
  &&= \frac{t^2}{B} 
  \left(|F_2| + |F_4| + |F_5| - |E_2| + |E_3|
  - 2|H| - 2|D_4| - 2|G_2| + 2|G_3| \right).
\end{eqnarray}
The sign of each contribution is the sign displayed next to it on the
corresponding diagram inverted if this diagram is from a 2-way
intersection of some sets (this is true for all sets but $D_4$).
The overall factor of two is due to the time-reversal symmetry and,
in case of the sets $F$ and $E$, it is cancelled by the diagram
symmetry factor of $1/2$, see Fig.~\ref{fig:all3}.

When evaluating the contributions we use the following rules of
thumb: (1) the length of the tangential intersections, denoted in the
diagrams by letters $a$ and/or $b$, can be set to 1; (2) the
restrictions can be ignored (see \ref{sec:app1} for a
discussion of such approximations).  We also notice that all orbits belonging
to the above diagrams give positive contribution (i.e. no more unitary
factors left), which shows we are on the right track.

\begin{figure}
  \def\fscale{0.8}
  \input{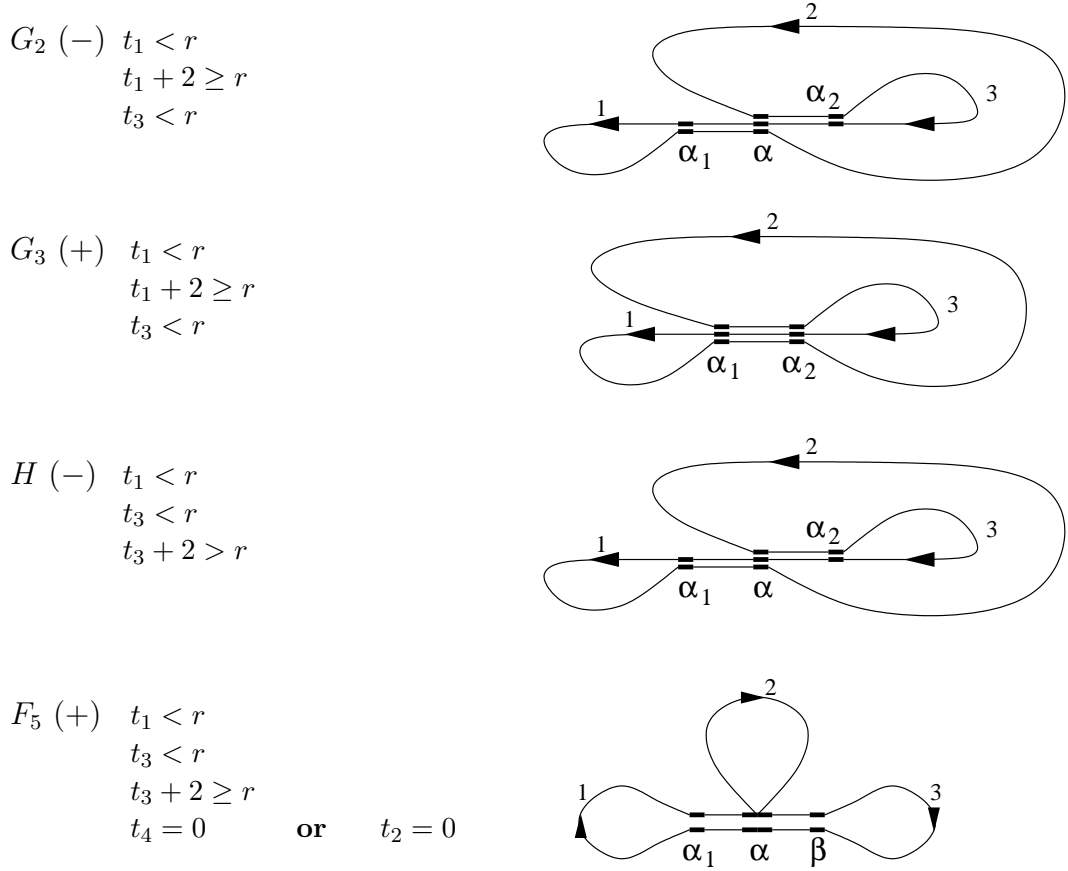}
  \caption{First three of the sets giving non-zero contributions}
  \label{fig:sets_G2_G3}
  \label{fig:sets_H_F5}
\end{figure}

We begin by showing that the contributions of the sets $G_2$ and
$G_3$ cancel each other.  On Fig.~\ref{fig:sets_G2_G3} we introduced some
extra notation necessary for the calculation.  The contribution of the
set $G_2$ can be written as
\begin{equation}
  \label{eq:contrib_G2}
  |G_2| = 2 \sum_{t_3<r} \sum_{\alpha_i}
   P_{(\alpha,\alpha_1)\ra(\alpha_1,\alpha)}^{t_1+1} 
   P_{(\alpha_1,\alpha)\ra(\alpha,\alpha_2)}^{t_2+1}
   P_{(\alpha,\alpha_2)\ra(\alpha_2,\alpha)}^{t_3+1}
   P_{(\alpha_2,\alpha)\ra(\alpha,\alpha_1)}^{1},
\end{equation}
where the summation is taken over $\alpha$s such that there exist
bonds $(\alpha,\alpha_1)$ and $(\alpha,\alpha_2)$ and over the ranges
of $t_i$s outlined on the diagram.  The two possible choices of $t_1$ are
expressed as the factor of $2$ and thus, due to the sum $t_1+t_2+t_3+4$
being fixed, the sum is essentially over $t_3<r$.  The arcs $1$ and
$2$ are (relatively) long, therefore the corresponding transition
probabilities can be approximated by their ergodic limit of $1/B$.  We
further sum over $\alpha_1$ to obtain
\begin{equation}
  \label{eq:contrib_G2_eval}
  \fl |G_2| = \frac{2}{B^2} \sum_{t_3<r} \sum_{\alpha_i}
   P_{(\alpha,\alpha_2)\ra(\alpha_2,\alpha)}^{t_3+1}
   P_{(\alpha_2,\alpha)\ra(\alpha,\alpha_1)}^{1} 
   = \frac{2}{B^2} \sum_{t_3<r} \sum_{\alpha, \alpha_2}
   P_{(\alpha,\alpha_2)\ra(\alpha_2,\alpha)}^{t_3+1}
\end{equation}

We evaluate the contribution of $G_3$ is a similar way,
\begin{eqnarray}
  \label{eq:contrib_G3}
  \fl |G_3| = 2 \sum_{t_3<r} \sum_{\alpha_1, \alpha_2}
   P_{(\alpha_2,\alpha_1)\ra(\alpha_1,\alpha_2)}^{t_1+1} 
   P_{(\alpha_1,\alpha_2)\ra(\alpha_1,\alpha_2)}^{t_2+1}
   P_{(\alpha_1,\alpha_2)\ra(\alpha_2,\alpha_1)}^{t_3+1} 
   \nonumber\\
   = \frac{2}{B^2} \sum_{t_3<r} \sum_{\alpha_1, \alpha_2}
   P_{(\alpha_1,\alpha_2)\ra(\alpha_2,\alpha_1)}^{t_3+1}
\end{eqnarray}
and notice that the final expressions for $|G_2|$ and $|G_3|$ are
identical and thus cancel each other.

The sets $H$ and $F_5$, Fig.~\ref{fig:sets_H_F5}, on the other hand,
are identical themselves.  The factor of 2 before the contribution of
$H$ is compensated by the factor of 2 in the diagram $F_5$ itself.
Thus the total contribution $|F_5| - 2|H| = 0$.

\begin{figure}
  \def\fscale{0.8}
  \input{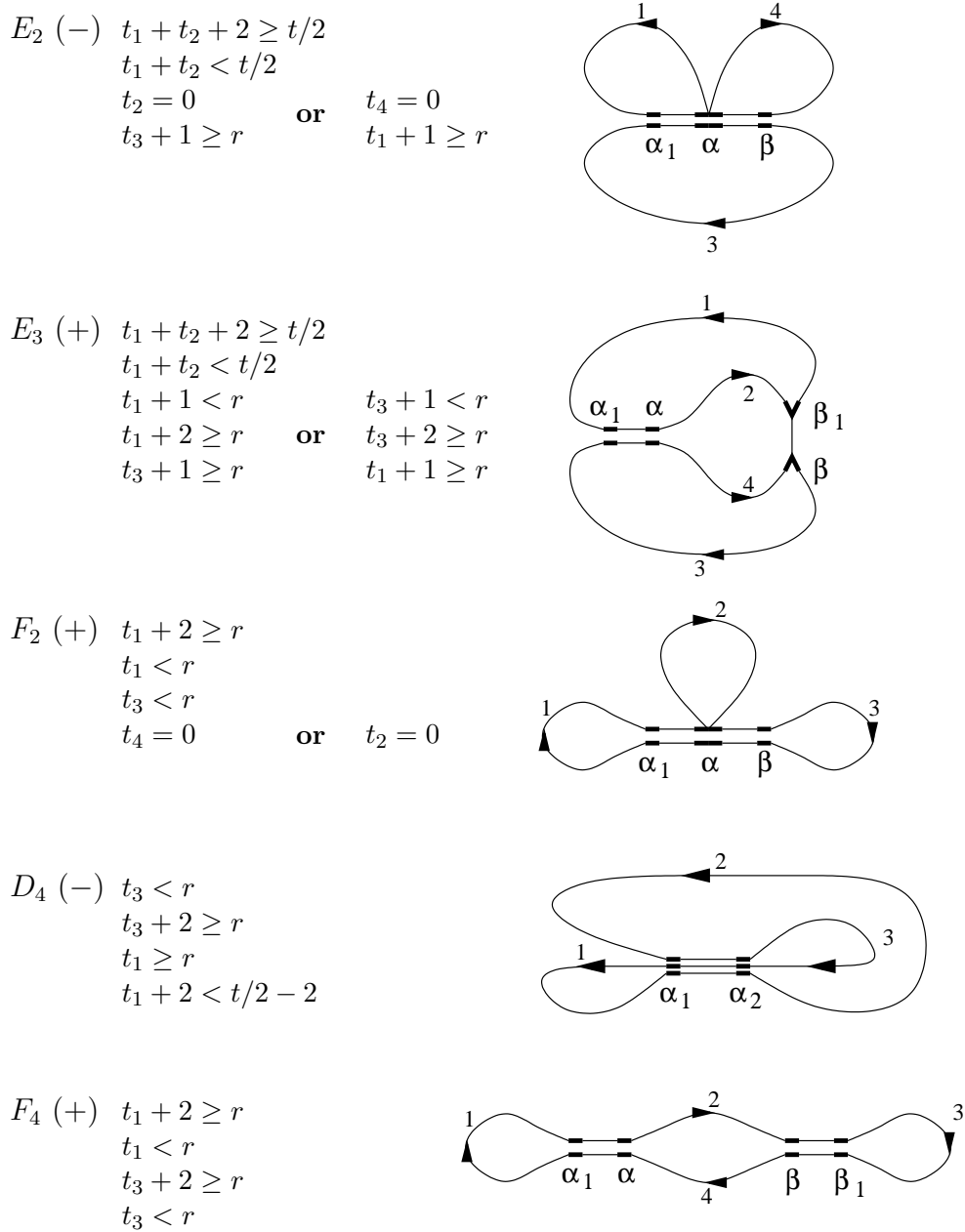}
  \caption{Sets $E_2$, $E_3$, $F_2$, $D_4$ and $F_4$ giving non-zero
    contributions.  The diagram of $E_2$ was ``symmetrised'' to
    highlight it's similarity with $F_2$}
  \label{fig:sets_E2_E3_F2}
  \label{fig:sets_D4}
  \label{fig:sets_F4}
\end{figure}

The contributions of the sets $E_2$, $E_3$ and $F_2$ are best grouped
together.  We start with $E_3$ (evaluating only one of the two cases
and multiplying it by 2), Fig.~\ref{fig:sets_E2_E3_F2}.  To simplify
the notation when dealing with numbers of order $t$, such as $t/2$ and
$r$, we ignore the constant corrections as they do not influence the
limiting behaviour.  The constraints imply $t_1 = r-2$ and $t_1+t_2 =
t/2 \pm \mathrm{const}$, therefore the summation is over $t_3$ (or
$t_4$).  The summation is limited by the constraint $t_3 > r$ on one
side and by $t_3 < t - t_1 - t_2 \approx t/2$ on the other.
\begin{eqnarray}
  \label{eq:contrib_E3}
  \fl |E_3| = 4 \sum_{r < t_3 < \frac{t}2}
  \sum_{\alpha, \alpha_1, \beta, \beta_1}
  P_{(\beta,\beta_1)\ra(\alpha_1,\alpha)}^{t_1+1} 
  P_{(\alpha_1,\alpha)\ra(\beta_1,\beta)}^{t_2+1}
  P_{(\beta_1,\beta)\ra(\alpha_1,\alpha)}^{t_3+1}
  P_{(\alpha_1,\alpha)\ra(\beta,\beta_1)}^{t_4+1} 
  \nonumber\\
  \lo= \frac{4}{B^3} \sum_{r < t_3 < \frac{t}2}
  \sum_{\alpha, \alpha_1, \beta, \beta_1}
  P_{(\alpha_1,\alpha)\ra(\beta,\beta_1)}^{t_4+1}
  = \frac{4}{B^3} \sum_{r < t_3 < \frac{t}2}
  \sum_{\alpha, \alpha_1} 1 
  = \frac{4}{B^2} \left(\frac{t}2 - r\right),
\end{eqnarray}
where to perform the summation over all possible bonds
$(\beta,\beta_1)$ we invoke the probability conservation.  The
summation over all possible bonds $(\alpha, \alpha_1)$ is just the
number of the bonds.

For $E_2$ the situation is very similar, but we choose to sum over
$t_4$,
\begin{eqnarray}
  \label{eq:contrib_E2}
  \fl  |E_2| = 4 \sum_{t_4 < \frac{t}2 - r}
  \sum_{\alpha, \alpha_1, \beta}
  P_{(\beta,\alpha)\ra(\alpha_1,\alpha)}^{t_1+1} 
  P_{(\alpha,\beta)\ra(\alpha_1,\alpha)}^{t_3+1}
  P_{(\alpha_1,\alpha)\ra(\beta,\alpha)}^{t_4+1} 
  P_{(\alpha_1,\alpha)\ra(\alpha,\beta)}^{1} 
  \nonumber\\
  \lo= \frac{4}{B^2} \sum_{t_4 < \frac{t}2 - r}
  \sum_{\alpha, \alpha_1, \beta}
  P_{(\alpha_1,\alpha)\ra(\beta,\alpha)}^{t_4+1}
  P_{(\alpha_1,\alpha)\ra(\alpha,\beta)}^{1}
  \nonumber\\
  \lo= \frac{4}{B^2} \sum_{t_4 < \frac{t}2 - r}
  \sum_{\alpha, \alpha_1}
  P_{(\alpha_1,\alpha)\ra(\alpha,\alpha_1)}^{t_4+2},
\end{eqnarray}
where, to get to the last line, we used the symmetry of the
time-reversal invariance, $P_{(\alpha_1,\alpha)\ra(\alpha,\beta)} =
P_{(\beta,\alpha)\ra(\alpha,\alpha_1)}$, and the Markov property of
the probabilities.  Unfortunately, the contribution of $E_2$ depends
on the particular structure of short orbits of the graph.  Hopefully, the
other contributions will help:
\begin{eqnarray}
  \label{eq:contrib_F2}
  \fl  |F_2| = 4 \sum_{t_3 < r}
  \sum_{\alpha, \alpha_1, \beta}
  P_{(\alpha,\alpha_1)\ra(\alpha_1,\alpha)}^{t_1+1} 
  P_{(\alpha_1,\alpha)\ra(\alpha,\beta)}^{t_2+1}
  P_{(\alpha,\beta)\ra(\beta,\alpha)}^{t_3+1} 
  P_{(\beta,\alpha)\ra(\alpha,\alpha_1)}^{1} 
  \nonumber\\
  \lo= \frac{4}{B^2} \sum_{t_3 < r}
  \sum_{\alpha, \alpha_1, \beta}
  P_{(\alpha,\beta)\ra(\beta,\alpha)}^{t_3+1} 
  P_{(\beta,\alpha)\ra(\alpha,\alpha_1)}^{1} 
  = \frac{4}{B^2} \sum_{t_3 < r}
  \sum_{\alpha, \beta}
  P_{(\alpha,\beta)\ra(\beta,\alpha)}^{t_3+1},
\end{eqnarray}
where to get the final result we summed over $\alpha_1$ and invoked
the probability conservation.  We notice that the sums in
(\ref{eq:contrib_F2}) and (\ref{eq:contrib_E2}) are identical, up to a
variable change, apart from the upper limit of the sum.  Assuming,
without loss of generality, that $r > \frac{t}2 - r$, we obtain
\begin{equation}
  \label{eq:contr_E2_F2}
  |F_2| - |E_2| = \frac{4}{B^2} \sum_{\frac{t}2 - r \leq t_3 < r}
  \sum_{\alpha, \beta}  P_{(\alpha,\beta)\ra(\beta,\alpha)}^{t_3+1},
\end{equation}
but now the sum is over large values of $t_3$, thus
\begin{equation}
  \label{eq:contr_E2_F2_eval}
  |F_2| - |E_2| = \frac{4}{B^3} \sum_{\frac{t}2 - r \leq t_3 < r}
  \sum_{\alpha, \beta} 1 = \frac{4}{B^2} \left(r - \left(\frac{t}2 -
  r\right)\right). 
\end{equation}
Bringing it together with $|E_3|$, we obtain
\begin{equation}
  \label{eq:contrib_E2_E3_F2}
  |F_2| - |E_2| + |E_3| = \frac{4r}{B^2}
\end{equation}

The next set to evaluate is $D_4$, Fig.~\ref{fig:sets_D4}.  All
arcs are long in this case, leading to the contribution
\begin{eqnarray}
  \label{eq:contrib_D4}
  |D_4| &=& 2 \sum_{r\leq t_1 < \frac{t}2}
  \sum_{\alpha_1, \alpha_2}
  P_{(\alpha_2,\alpha_1)\ra(\alpha_1,\alpha_2)}^{t_1+1} 
  P_{(\alpha_1,\alpha_2)\ra(\alpha_1,\alpha_2)}^{t_2+1}
  P_{(\alpha_1,\alpha_2)\ra(\alpha_2,\alpha_1)}^{t_3+1} 
  \nonumber\\
  &=& \frac{2}{B^3} \sum_{r\leq t_1 < \frac{t}2}
  \sum_{\alpha_1, \alpha_2} 1
  = \frac{2}{B^2} \left(\frac{t}2 - r\right).
\end{eqnarray}

In the final contribution, coming from the set $F_4$ of
Fig.~\ref{fig:sets_F4}, the summation is over $t_2+t_4\approx t-2r$.
The arcs $1$ and $3$ are both long and of fixed length $r-1$ or $r-2$.
This freedom of choice gives rise to the $4$ factor in front of the
sum,
\begin{equation}
  \label{eq:contrib_F4}
  |F_4| = \frac{4}{B^2} \sum_{t_2+t_4 = t-2r}
  \sum_{\alpha, \alpha_1, \beta, \beta_1}
  P_{(\alpha_1,\alpha)\ra(\beta,\beta_1)}^{t_2+1}
  P_{(\beta_1,\beta)\ra(\alpha,\alpha_1)}^{t_4+1},
\end{equation}
where the terms corresponding to the arcs $1$ and $3$, have already
being approximated by $1/B$.  At least one of the arcs, $2$ or $4$ is
long.  Assuming, without loss of generality, that $2$ is long, we have
for the inner sum,
\begin{equation}
  \label{eq:contrib_F4_inner}
  \fl \sum_{\alpha, \alpha_1, \beta, \beta_1}
  P_{(\alpha_1,\alpha)\ra(\beta,\beta_1)}^{t_2+1}
  P_{(\beta_1,\beta)\ra(\alpha,\alpha_1)}^{t_4+1} 
  = \frac1B \sum_{\alpha, \alpha_1, \beta, \beta_1}
  P_{(\beta_1,\beta)\ra(\alpha,\alpha_1)}^{t_4+1} 
  = \frac 1B \sum_{\beta, \beta_1} 1 = 1.
\end{equation}
Thus the result for the set $F_4$ is
\begin{equation}
  \label{contrib_F4_res}
  |F_4| = \frac{4}{B^2} \sum_{t_2+t_4 = t-2r} 1 = \frac{4(t-r-r)}{B^2}.
\end{equation}

The overall result is
\begin{eqnarray}
  \label{eq:significant_sets_res}
  \fl 2(K_{\rm TR3a}(\tau) + K_{\rm TR3b}(\tau) + K_{\rm TR3c}(\tau))
  = \frac{t^2}{B} 
  \left(|F_4| + |F_2| - |E_2| + |E_3|  - 2|D_4|\right)
  \nonumber\\
  \lo= \frac{t^2}{B} \left( 
    \frac{4}{B^2}(t-r-r) + \frac{4}{B^2}r 
    - \frac{4}{B^2}\left(\frac{t}2 - r\right)
  \right) = \frac{2t^3}{B^3} \to 2\tau^3.
\end{eqnarray}
Taking into account that $K_{\rm NTR3a}(\tau) + K_{\rm NTR3b}(\tau) =
0$ \cite{BSWpre}, we obtain the final result $K_{\rm TR3}(\tau) =
2\tau^3$, QED.

\section{Conclusions}
The results of this manuscript complement those of \cite{BSWpre} to
form the derivation of the form factor of {\em generic} quantum graphs
to the third order.

While the derivation presented above is quite technical, the
underlying idea is beautiful in its simplicity: the set of all orbits
can be partitioned in such a way that the contribution of each
partition set can be easily shown to be zero.  And this can be done
without evaluating the contributions of individual orbits!  The
nonzero ``correct'' result then arises from the intersections of the
partition sets.  The orbits in such intersections have generic
properties: most of their arcs are long, making direct evaluation possible.

We presented one of the possible partitions and obtained the expected RMT
result.  The partition is based upon three basic diagrams, each being
separately partitioned.  It is very possible that there exists a
``unified'' diagram with a simple partition.  If such diagram is
found, it might make possible the derivation to all orders.  Alas, it
has so far evaded all attempts to find it.

The derivation is done for sequences of graphs satisfying two
conditions: (a) graphs must be ``uniformly ergodic'', as expressed by
Eq.~(\ref{eq:ergodic-tau}) and (b) the contribution of long
self-retracing orbits must be negligible,
Eq.~(\ref{eq:matrixR_decays}).  These conditions exclude such cases as
Neumann star graphs, which are known to possess statistics different
from RMT predictions.  The families of graphs numerically found to
satisfy the RMT hypothesis, such as Neumann complete graphs and
Fourier star graphs, also satisfy the above conditions.

\section{Acknowledgements}
The idea of the evaluation of the TR2 diagram, as it is done in
Section~\ref{sec:TR2}, was born in a discussion with Holger Schanz
during the author's visit to Max-Planck-Institut, G{\"o}ttingen,
Germany.  This idea later become the foundation for this entire work.
The author is extremely grateful to Robert Whitney who proofread the
manuscript, pointed out numerous mistakes and proposed
simplifications.  The manuscript was further improved
following extremely helpful remarks of the referee.  A part of the
research was performed while the author was working at the Weizmann
Institute of Science, Rehovot, Israel, supported by the Israel Science
Foundation, a Minerva grant, and the Minerva Center for Nonlinear
Physics.  This, and the encouragement provided by U. Smilansky, is
greatly appreciated.

\appendix
\section{Setting $a=1$ in TR2 diagram}
\label{sec:app1}

\begin{figure}[t]
  \input{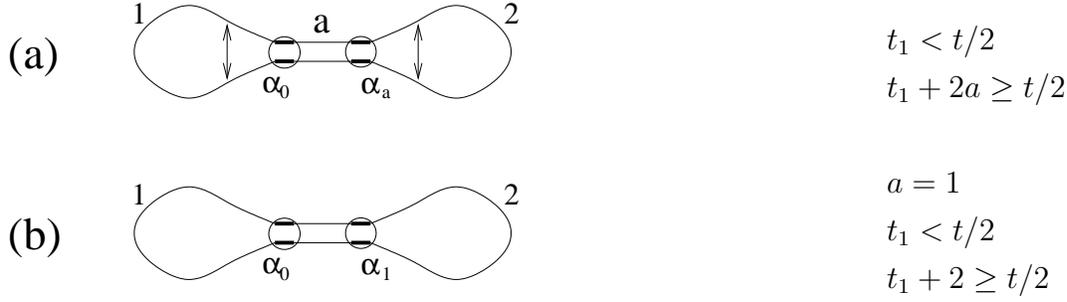}
  \caption{In the first set, the tangential intersection must be of 
    length $a>0$.  In the second set, the length of the intersection
    is not explicitly  specified.  It turns out that both sets contain
    exactly the same  orbits.}
  \label{fig:app1}
\end{figure}

When we evaluated the contribution of the TR2 orbits, we dropped the
constraints and set the length of tangential intersection $a$ to 1
(Fig.~\ref{fig:app1}).  Here we discuss various aspects of this
approximation.

\begin{figure}[t]
  \def\fscale{1.0}
  \centerline{\includegraphics[scale=\fscale]{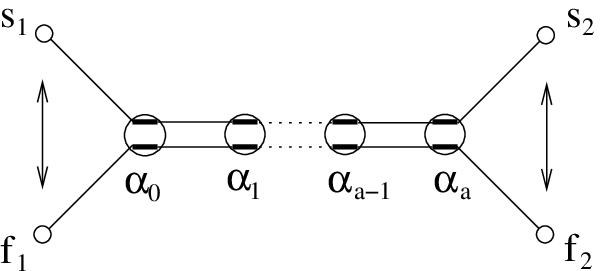}}
  \caption{Detailed view of the tangential intersection of length $a$, 
    together with the adjacent vertices $s_i$ and $f_i$.}
  \label{fig:app2}
\end{figure}

For a given value of $a$, there are $2a$ choices of the length $t_1$
satisfying the inequalities on part (a) of Fig.~\ref{fig:app1}.  For
small $a$, the length of both arcs $(\alpha_0, s_1) \to (f_1,
\alpha_0)$ and $(\alpha_a, s_2) \to (f_2, \alpha_a)$ are long and
their contributions can be approximated by $1/B$.  Thus the
contribution of the TR2 diagram is
\begin{equation}
  \label{eq:TR2_contrib_full}
  \frac{t^2}{B^3} \sum_{a=1}^{a'} 2a T_a,
\end{equation}
where $T_a$ is the contribution of the structure depicted on
Fig.~\ref{fig:app2}, the tangential intersection of the length $a$.
This contribution can be written as 
\begin{equation}
  \label{eq:T_a}
  \fl T_a = \sum |\sigma_{s_1, \alpha_1}|^2 |\sigma_{f_1, \alpha_1}|^2
  \prod_{k=0}^{a-2} |\sigma_{\alpha_k, \alpha_{k+2}}|^4  
  |\sigma_{s_2, \alpha_{a-2}}|^2 |\sigma_{f_2, \alpha_{a-2}}|^2
  \left(1 - \delta_{s_1, f_1}\right) \left(1 - \delta_{s_2, f_2}\right),
\end{equation}
where the summation is taken over all indices $s_i$, $f_i$
and $\alpha_k$.  Our aim is to show that the sum of $2aT_a$ is equal
to the corresponding contribution from part (b) of
Fig.~\ref{fig:app1}, given by $\displaystyle{2 \sum_{\alpha_0, \alpha_1} 1}$.
Opening up the brackets in (\ref{eq:T_a}) and performing summation,
where possible, over $s_i$ and $f_i$, we get
\begin{eqnarray}
  \label{eq:T_a_expanded}
  \fl T_a = \sum \prod_{k=0}^{a-2} |\sigma_{\alpha_k, \alpha_{k+2}}|^4  
  - \sum |\sigma_{s_1, \alpha_1}|^4
  \prod_{k=0}^{a-2} |\sigma_{\alpha_k, \alpha_{k+2}}|^4
  - \sum \prod_{k=0}^{a-2} |\sigma_{\alpha_k, \alpha_{k+2}}|^4  
  |\sigma_{s_2, \alpha_{a-2}}|^4 \\
  + \sum |\sigma_{s_1, \alpha_1}|^4
  \prod_{k=0}^{a-2} |\sigma_{\alpha_k, \alpha_{k+2}}|^4  
  |\sigma_{s_2, \alpha_{a-2}}|^4 \\
  \lo= R_a - 2R_{a+1} + R_{a+2},
\end{eqnarray}
where $R_a$ is
\begin{equation}
  \label{eq:R_a_def}
  R_a = \sum \prod_{k=0}^{a-2} |\sigma_{\alpha_k, \alpha_{k+2}}|^4.
\end{equation}
The sum in (\ref{eq:TR2_contrib_full}) now folds to
\begin{equation}
  \label{eq:TR2_contrib_folds}
  \sum_{a=1}^{a'} 2a T_a = 2R_0 + 2a'R_{a'+2} - (2a'+2)R_{a'+1}.
\end{equation}
Since $R_0$ is nothing else than $\displaystyle{\sum_{\alpha_0,
    \alpha_1} 1}$, to finish the proof we need to ensure that (i) the
remainder of (\ref{eq:TR2_contrib_folds}) is negligible and (ii) we
can ignore orbits with $a > a'$.  To this end we take $a'$ to be of
order $t$ and refer to condition~(\ref{eq:matrixR_decays}) which is
obviously sufficient.

\section*{References}
\bibliographystyle{unsrt}\bibliography{diagrams}

\begin{thebibliography}{10}

\bibitem{BGS84}
O.~Bohigas, M.~J. Giannoni, and C.~Schmit.
\newblock {\em Phys. Rev. Lett.}, 52:1--4, 1984.

\bibitem{Haake}
F.~Haake.
\newblock {\em Quantum Signatures of Chaos}.
\newblock Springer, Berlin, 2000.

\bibitem{Ber85}
M.~V. Berry.
\newblock {\em Proc. R. Soc. Lond. A}, 400:229--251, 1985.

\bibitem{SR01}
M.~Sieber and K.~Richter.
\newblock {\em Physica Scripta}, T90:128, 2001.

\bibitem{S02}
M.~Sieber.
\newblock {\em J.~Phys.~A} 35:L613-L619, 2002 (also nlin.CD/0209016, 2002).

\bibitem{Gut71}
M.~C. Gutzwiller.
\newblock {\em J. Math. Phys.}, 12:343--358, 1971.

\bibitem{KS97}
T.~Kottos and U.~Smilansky.
\newblock {\em Phys.~Rev.~Lett.}, 79:4794--4797, 1997.

\bibitem{KS99}
T.~Kottos and U.~Smilansky.
\newblock {\em Ann.~Phys.}, 274:76--124, 1999.

\bibitem{BSW02}
G.~Berkolaiko, H.~Schanz, and R.~S. Whitney.
\newblock {\em Phys. Rev. Lett.}, 88:104101, 2002.

\bibitem{BSWpre}
G.~Berkolaiko, H.~Schanz, and R.~S. Whitney.
\newblock {\em Preprint} nlin.CD/0205014, 2002.


\bibitem{Mehta}
M.~L. Mehta.
\newblock Academic Press, New York, 1991.

\bibitem{BG00}
F.~Barra and P.~Gaspard.
\newblock {\em J.~Stat.~Phys.}, 101:283--319, 2000.

\bibitem{Grim}
G.R.~Grimmett and D.R.~Stirzaker.
\newblock {\em Probability and Random Processes}.
\newblock Claredon Press, Oxford, 1992.

\bibitem{Tan01}
G.~Tanner.
\newblock {\em J.~Phys.~A}, 34:8485, 2001.

\bibitem{BK99}
G.~Berkolaiko and J.P.~Keating.
\newblock {\em J.~Phys.~A}, 32:7827--7841, 1999.

\end{thebibliography}

\end{document}